\documentclass[11pt,a4paper]{article}
\usepackage{amsmath}
\usepackage{cite}
\usepackage{graphicx}
\usepackage{amsfonts}
\usepackage{amssymb}

\def\bt{
\beta}

\def\bt'{\beta'
}
\begin{document}
\bigskip
\hfill\hbox{SPhT-T03/052} \vspace{2cm}

\begin{center}
{\Large \textbf{Massless N=1 Super Sinh-Gordon:\\ \vspace{0.5cm}  Form Factors approach}} \\
\vspace{1.2cm} {\Large B\'en\'edicte Ponsot \footnote{\textsf
{ponsot@spht.saclay.cea.fr}}} \\
\vspace{0.7cm} {\it  Service de Physique Th\'eorique, Commissariat
\`a l'\'energie
atomique,\\
L'Orme des Merisiers,
 F-91191 Gif sur Yvette, France.}\\
\vspace{1.2cm}
\end{center}

\begin{abstract}
The $N=1$ Super Sinh-Gordon model with spontaneously broken
supersymmetry is considered. Explicit expressions for form-factors
of the trace of the stress energy tensor $\Theta$, the energy
operator $\epsilon$, as well as the order and disorder operators $\sigma$ and $\mu$ are proposed.
\end{abstract}
\begin{center}
\end{center}

\vspace{1cm}
\section{Introduction}
The SShG model can be considered as a perturbed super Liouville
field theory, which lagrangian density is given by
$$
\mathcal{L}=\frac{1}{8\pi}(\partial_a
\phi)^2-\frac{1}{2\pi}(\bar{\psi}\partial
\bar{\psi}+\psi\bar{\partial} \psi) +i\mu b^2\psi \bar{\psi}
e^{b\phi}+\frac{\pi \mu^2 b^2}{2}e^{2b\phi}.
$$
with the background charge $Q=b+1/b$. This model is a CFT with
central charge
$$
c_{SL}=\frac{3}{2}(1+2Q^2).
$$
The super Sinh-Gordon model is 1+1 dimensional integrable quantum
field theory with $N=1$ supersymmetry. We consider the Lagrangian
$$
\mathcal{L}=\frac{1}{8\pi}(\partial_a
\phi)^2-\frac{1}{2\pi}(\bar{\psi}\partial
\bar{\psi}+\psi\bar{\partial} \psi) +2i\mu b^2\psi \bar{\psi}
\sinh b\phi+2\pi \mu^2 b^2\cosh^2 b\phi.
$$
In this model the supersymmetry is spontaneously broken
\cite{AKRZ}: the bosonic field becomes massive, but the Majorana fermion
stays massless and plays the role of Goldstino. In the IR limit, the
 effective theory for the Goldstino is to the lowest order
 the Volkov-Akulov lagrangian \cite{VA}
\begin{eqnarray}
\mathcal{L}_{IR}=(\bar{\psi}\partial \bar{\psi}+\psi\bar{\partial}
\psi)-\frac{4}{M^2}(\psi\partial \psi)(\bar{\psi}\bar{\partial}
\bar{\psi})+\cdots
\label{effectif}
\end{eqnarray}
where supersymmetry is realized non linearly. The irrelevant
operator along which the Super Liouville theory flows into Ising
is the product of stress-energy tensor $T\bar{T}=(\psi\partial \psi)(\bar{\psi}\bar{\partial}
\bar{\psi})$, which is the
lowest dimension non derivative operator allowed by the
symmetries. The dots include higher dimensional irrelevant
operators.\\
The scattering in the left-left and right-right subchannels is
trivial, but not in the right-left channel. The following
scattering matrices were proposed in \cite{AKRZ}
$$
S_{RR}(\theta)=S_{LL}(\theta')=-1,\; S_{RL}(\theta - \theta')=
-\frac{\sinh(\theta-\theta')-i\sin\pi\nu}{\sinh(\theta-\theta')+i\sin\pi\nu}
,\quad \nu\equiv b/Q.
$$
For the right and left movers, the energy momentum is parametrized in
terms of the rapidity variables $\theta$ and $\theta'$ by
$p^0=p^1=\frac{M}{2}e^{\theta}$ (and
$p^0=-p^1=\frac{M}{2}e^{-\theta'}$). The mass scale of the theory
$M^{-2}$ is equal to $2\sin \pi \nu$. The form factors\footnote{We
refer the reader to \cite{DMS} for a discussion on form factors in
massless QFT.}
$F_{r,l}(\theta_1,\theta_2,\ldots,\theta_r;\theta'_1,\theta'_2,\ldots,\theta'_l)$
are defined to be matrix elements of an operator between the
vacuum and a set of asymptotics states. The form factor bootstrap
approach \cite{KW,BKW,S} (developed originally for massive
theories, but that turned out to be also an effective tool for massless theories \cite{DMS,MS}) leads to a system of
linear functional relations for the matrix elements $F_{r,l}$; let
us introduce the minimal form factors which have neither poles nor
zeros in the strip $0<\Im m\theta<\pi$ and which are solutions of
the equations $f_{\alpha_1 \alpha_2}(\theta)=f_{\alpha_1
\alpha_2}(\theta+2i\pi)S_{\alpha_1\alpha_2}(\theta),\;
\alpha_{i}=R,L$.\\
Then the general form factor is parametrized as follows:
\begin{eqnarray}
\lefteqn{F_{r,l}^{\alpha}(\theta_1,\theta_2,\ldots,\theta_r;\theta'_1,\theta'_2,\ldots,\theta'_l)=}
\nonumber \\
&&
 \prod_{1\leq i<j\leq r}f_{RR}(\theta_i -\theta_j)
\prod_{i=1}^{r}\prod_{j=1}^{l}f_{RL}(\theta_i -\theta'_j)
\prod_{1\leq i<j\leq l}f_{LL}(\theta'_i -\theta'_j)
Q_{r,l}(\theta_1,\theta_2,\ldots\theta_r;\theta'_1,\theta'_2,\ldots,\theta'_l),
\nonumber
\end{eqnarray}
and the function $Q_{r,l}$ depends on the operator considered. The
$RR$ and $LL$ scattering formally behave as in the massive case,
so annihilation poles occur in the $RR$ and $LL$ subchannel. This leads to the residue formula
\begin{eqnarray}
\lefteqn{\mathrm{Res}_{\theta_{12}=i\pi}F_{r,l}(\theta_1,\theta_2,\ldots,\theta_r;\theta'_1,\theta'_2,\ldots,\theta'_l)=}
\nonumber \\
&&
2F_{r-2,l}(\theta_3,\ldots,\theta_r;\theta'_1,\theta'_2,\ldots,\theta'_l)\left(1-\prod_{j=3}^{r}
S_{RR}(\theta_{2i})\prod_{k=1}^{l}S_{RL}(\theta_2-\theta'_k)\right),
\label{residue}
\end{eqnarray}
and a similar expression in the $LL$ subchannel.
 It
is important to note that these equations {\it do not} refer to
any specific operator.

\section{Expression for form factors}
 The minimal form factors read explicitly:
$$
f_{RR}(\theta)=\sinh\frac{\theta}{2}, \ \
f_{LL}(\theta')=\sinh\frac{\theta'}{2},
$$
and
$$
f_{RL}(\theta) = \frac{1}{2\cosh\frac{\theta}{2}}\ \ \exp
\int_{0}^{\infty}
\frac{dt}{t}\frac{\cosh(\frac{1}{2}-\nu)t-\cosh\frac{1}{2}t}{\sinh
t \cosh t/2}\cosh t\left(1-\frac{\theta}{i\pi}\right).
$$
The latter form factor has asymptotic behaviour when $\theta \to
-\infty$
\begin{eqnarray}
 f(\theta) \sim
e^{\theta/2}\left(1+\left(A + A'\theta\right) e^{\theta}+
\left(\frac{A^2}{2}+B+
AA'\theta+\frac{(A')^2\theta^2}{2}\right)e^{2\theta}\right).
\label{IR}
\end{eqnarray}
where $A=(1-2\nu)\cos\pi\nu -1 + 2i\sin\pi\nu, \;
A'=-\frac{2}{\pi}\sin\pi\nu, \; B=\frac{1}{2}(\cos 2\pi\nu -1)$.
The logarithmic contributions come from resonances.\\
The residue condition (\ref{residue}) written in terms of the
function $Q_{r,l}$ reads
\begin{eqnarray}
\lefteqn{\mathrm{Res}_{\theta_{12}=i\pi}Q_{r,l}(\theta_1,\theta_2,\ldots\theta_r;
\theta'_1,\theta'_2,\ldots,\theta'_l)
=Q_{r-2,l}(\theta_3,\ldots\theta_r;\theta'_1,\theta'_2,\ldots,\theta'_l)\times
(-)^{r-1}
(2i)^{l+r-1}\times} \nonumber \\
&& \times\prod_{j=3}^{r}\frac{1}{\sinh\theta_{2j}} \left(
\prod_{k=1}^{l}(\sinh (\theta_2-\theta'_k)+i\sin\pi \nu)
-(-1)^{r+l}\prod_{k=1}^{l}(\sinh (\theta_2-\theta'_k)-i\sin\pi
\nu) \right).
\label{residu2}
\end{eqnarray}
Let us introduce now the functions
\begin{eqnarray}
\phi(\theta_{ij}) \equiv
\frac{S_{RR}}{f_{RR}(\theta_{ij})f_{RR}(\theta_{ij}+i\pi)}
=\frac{2i}{\sinh \theta_{ij}}\; ,\quad \phi(\theta'_{ij}) \equiv
\frac{S_{LL}}{f_{LL}(\theta'_{ij})f_{LL}(\theta'_{ij}+i\pi)} =
\frac{2i}{\sinh \theta'_{ij}}\; . \nonumber
\end{eqnarray}
as well as
\begin{eqnarray}
\Phi(\theta_i-\theta'_j) &&\equiv
\frac{S_{RL}(\theta_i-\theta'_j)}{f_{RL}(\theta_i-\theta'_j)f_{RL}(\theta_i-\theta'_j+i\pi)}
= -2i\left(\sinh(\theta_i-\theta'_j)-i\sin \pi \nu\right)\; ,
\nonumber
\end{eqnarray}
and
\begin{eqnarray}
\tilde{\Phi}(\theta_i-\theta'_j) \equiv
\Phi(\theta_i-\theta'_j+i\pi)=2i\left(
\sinh(\theta_i-\theta'_j)+i\sin \pi \nu\right)\; . \nonumber
\end{eqnarray}
We assign odd $Z_2$-parity to both right and left-movers ($\psi_R
\to -\psi_R, \; \psi_L \to -\psi_L, \; \phi \to \phi)$  and even
(odd) parity to right (left) movers under duality transformations
($\psi_R \to \psi_R, \; \psi_L \to -\psi_L, \; \phi \to -\phi).$

\subsection{Neveu-Schwarz sector: trace of the stress-energy tensor}
The operator $\Theta$ has non zero matrix elements on (even,even)
number of particles. The first form factor is determined by using
the Lagrangian: $Q_{2,2}=-4\pi M^2$. We introduce the sets
$S=(1,\dots,2r)$ and $S'=(1,\dots,2l)$, and propose
\begin{eqnarray}
\lefteqn{Q_{2r,2l}(\theta_1,\theta_2,\ldots
\theta_{2r};\theta'_1,\theta'_2,\ldots,\theta'_{2l})=}\nonumber \\
&& -4\pi M^2\sum_{T \in S, \atop \#T=r-1}\sum_{T' \in S',\atop
\#T'=l-1}\prod_{i \in T, \atop k\in
\bar{T}}\phi(\theta_{ik})\prod_{i \in T', \atop  k\in
\bar{T}'}\phi(\theta'_{ik}) \prod_{i \in T, \atop k\in
\bar{T}'}\Phi(\theta_i-\theta'_k) \prod_{i \in T', \atop k\in
\bar{T}}\tilde{\Phi}(\theta_k-\theta'_i)\nonumber
\end{eqnarray}
where $T,\bar{T}$ are respectively subsets of $S$ and $\bar{S}$, the notation '$\#$' stands for 'number of elements', and by definition $\bar{T}=S\backslash T$, $\bar{T'}=S'\backslash T'$.\\
Let us show that this representation does indeed satisfy the
residue condition (\ref{residu2}): only two cases will contribute
to this computation, namely when $1 \in T, 2\in \bar{T}$ and $2
\in T, 1\in \bar{T}$. It amounts to evaluate the residue at
$\theta_{12}=i\pi$ of the quantity:
\begin{eqnarray}
  \lefteqn{\left[\phi(\theta_{12})\prod_{k\in \bar{T}-\{2\}}
      \phi(\theta_{1k})\prod_{i\in T-\{1\}} \phi(\theta_{i2})
      \prod_{k\in \bar{T}'}\Phi(\theta_1-\theta'_k) \prod_{i \in
	T'}\tilde{\Phi}(\theta_2-\theta'_i) \right.}
    \nonumber \\
    && \left. +\phi(\theta_{21})\prod_{k\in \bar{T}-\{1\}}
    \phi(\theta_{2k})\prod_{i\in T-\{2\}} \phi(\theta_{i1})
    \prod_{k\in \bar{T}'}\Phi(\theta_2-\theta'_k) \prod_{i \in
      T'}\tilde{\Phi}(\theta_1-\theta'_i)\right]\times \nonumber \\
  &&\times \;  -4\pi M^2\sum_{U \in S-\{1,2\}, \atop \#U=r-2}\sum_{T' \in S',\atop
    \#T'=l-1}\prod_{i \in U, \atop k\in
    \bar{U}}\phi(\theta_{ik})\prod_{i \in T', \atop  k\in
    \bar{T}'}\phi(\theta'_{ik}) \prod_{i \in U, \atop k\in
    \bar{T}'}\Phi(\theta_i-\theta'_k) \prod_{i \in T', \atop k\in
    \bar{U}}\tilde{\Phi}(\theta_k-\theta'_i).\nonumber
\end{eqnarray}
The last line is nothing but $Q_{2r-2,2l}(\theta_3,\theta_4,\ldots
\theta_{2r};\theta'_1,\theta'_2,\ldots,\theta'_{2l})$; the
evaluation of the residue at $\theta_{12}=i\pi$ of the term into
brackets gives explicitly
\begin{eqnarray}
  (2i)^{2r+2l-1}\prod_{j=3}^{2r}\frac{1}{\sinh\theta_{2j}}
  \left[(-1)^{2r-2}\times  \prod_{k=1}^{2l}(\sinh
    (\theta_2-\theta'_k)-i\sin\pi \nu)
    -(-1)^{2l}\prod_{k=1}^{2l}(\sinh (\theta_2-\theta'_k)+i\sin\pi
    \nu) \right],\nonumber
\end{eqnarray}
and equation (\ref{residu2}) is satisfied.\\
As a remark, we would
like to note that the leading infrared behaviour of $F_{2,2}$ is
given by $T\bar{T}$, which defines the direction of the flow in
the IR region. To determine the subleading IR terms that appear in
the expansion (\ref{effectif}), one uses the asymptotic
development for $f_{RL}$ given by equation (\ref{IR}). For example
(up to the logarithmic terms):
\begin{eqnarray}
  &&\lefteqn{f_{RL}(\theta_1-\theta'_1)f_{RL}(\theta_1-\theta'_2)f_{RL}(\theta_2-\theta'_1)f_{RL}(\theta_2-\theta'_2)
    \sim e^{\theta_1+\theta_2-\theta'_1-\theta'_2}\times} \nonumber \\
  &&\left[1+Ae^{\theta_1-\theta'_1}+\left(\frac{A^2}{2}+B\right)e^{2\theta_1-2\theta'_1}\right]\times
  \left[1+Ae^{\theta_1-\theta'_2}+\left(\frac{A^2}{2}+B\right)e^{2\theta_1-2\theta'_2}\right]\times
  \nonumber \\
  &&
  \left[1+Ae^{\theta_2-\theta'_1}+\left(\frac{A^2}{2}+B\right)e^{2\theta_2-2\theta'_1}\right]\times
  \left[1+Ae^{\theta_2-\theta'_2}+\left(\frac{A^2}{2}+B\right)e^{2\theta_2-2\theta'_2}\right]\;
  . \nonumber
\end{eqnarray}
The terms into brackets give
\begin{eqnarray}
  &&1+A(e^{\theta_1}+e^{\theta_2})(e^{-\theta'_1}+e^{-\theta'_2})+
  \left(\frac{A^2}{2}+B\right)(e^{2\theta_1-2\theta'_1}+e^{2\theta_1-2\theta'_2}+e^{2\theta_2-2\theta'_1}
  +e^{2\theta_2-2\theta'_2})
  \nonumber\\
  &&+A^2(e^{2\theta_1-\theta'_1-\theta'_2}+
  e^{\theta_1+\theta_2-2\theta'_1}+e^{\theta_1+\theta_2-2\theta'_2}+e^{2\theta_2-\theta'_1-\theta'_2}+
  2e^{\theta_1+\theta_2-\theta'_1-\theta'_2})
  +\cdots\nonumber\\
  &&=
  1+\frac{A}{M^2}L_{-1}\bar{L}_{-1}+\frac{A^2}{2M^4}L^2_{-1}\bar{L}^2_{-1}+
  \frac{B}{M^4}L_{-2}\bar{L}_{-2}+\cdots\nonumber
\end{eqnarray}
where $L_{-1}=e^{\theta_1}+e^{\theta_2}$ and
$L_{-2}=e^{2\theta_1}+e^{2\theta_2}$. So the next irrelevant
operator appearing in (\ref{effectif}) is $T^2\bar{T}^2$ (up to
derivatives).

\subsubsection{Form factors of the energy operator $\epsilon$.}
The number of left movers and right movers is odd. Let
$S=(1,\dots,2r+1),S'=(1,\dots,2l+1)$. The lowest form factor is
$Q_{1,1}=1$. We propose
\begin{eqnarray}
  \lefteqn{Q_{2r+1,2l+1}(\theta_1,\theta_2,\ldots
    \theta_{2r+1};\theta'_1,\theta'_2,\ldots,\theta'_{2l+1})=}
  \nonumber \\
  && \sum_{T \in S, \atop \#T=r}\sum_{T' \in S',\atop
    \#T'=l}\prod_{i \in T, \atop k\in
    \bar{T}}\phi(\theta_{ik})\prod_{i \in T', \atop k\in
    \bar{T}'}\phi(\theta'_{ik}) \prod_{i \in T, \atop k\in
    \bar{T}'}\Phi(\theta_i-\theta'_k) \prod_{i \in T', \atop k\in
    \bar{T}}\tilde{\Phi}(\theta_k-\theta'_i)\nonumber.
\end{eqnarray}
The proof that this expression satisfies equation (\ref{residu2})
is the same as above.

\subsection{Ramond sector}
\subsubsection{Order operator $\sigma $.}
It has non vanishing matrix elements when the sum of left movers
and right movers is odd. Let $S=(1,\dots,2r+1),S'=(1,\dots,2l)$.
The lowest form factors are $Q_{1,0}=Q_{0,1}=1$.
We propose:
\begin{eqnarray}
  \lefteqn{Q_{2r+1,2l}(\theta_1,\theta_2,\ldots
    \theta_{2r+1};\theta'_1,\theta'_2,\ldots,\theta'_{2l})=}\nonumber \\
  && \sum_{T \in S, \atop \#T=r}\sum_{T' \in S', \atop
    \#T'=l}\prod_{i \in T, \atop k\in
    \bar{T}}\phi(\theta_{ik})\prod_{i \in T', \atop k\in
    \bar{T}'}\phi(\theta'_{ik}) \prod_{i \in T, \atop k\in
    \bar{T}'}\Phi(\theta_i-\theta'_k) \prod_{i \in T', \atop k\in
    \bar{T}}\tilde{\Phi}(\theta_k-\theta'_i)\nonumber
\end{eqnarray}

\subsubsection{Disorder operator $\mu$.}
It has non vanishing matrix elements when the sum of left and
right movers is even. As it is explained in \cite{YZ}, there is an
additional minus sign in front of the product of $S$ matrices in
the residue condition (\ref{residue}).
\begin{itemize}
\item the number of left and right movers are both even\\
  Let $S=(1,\dots,2r), S'=(1,\dots,2l)$ and the lowest form factor
  $Q_{0,0}=1$. We propose :
  \begin{eqnarray}
    \lefteqn{Q_{2r,2l}(\theta_1,\theta_2,\ldots
      \theta_{2r};\theta'_1,\theta'_2,\ldots,\theta'_{2l})=}\nonumber \\
    && (-i)^{r+l}\sum_{T \in S, \atop \#T=r}\sum_{T' \in S', \atop
      \#T'=l}\prod_{i \in T, \atop k\in
      \bar{T}}\phi(\theta_{ik})e^{\frac{1}{2}\sum\theta_{ki}}\prod_{i
      \in T',\atop  k\in \bar{T}'}\phi(\theta'_{ik})
    e^{\frac{1}{2}\sum\theta'_{ik}} \prod_{i \in T, \atop k\in
      \bar{T}'}\Phi(\theta_i-\theta'_k) \prod_{i \in T', \atop k\in
      \bar{T}}\tilde{\Phi}(\theta_k-\theta'_i)\nonumber
  \end{eqnarray}

\item{the number of left and right movers are both odd}\\
  Let $S=(1,\dots,2r+1),S'=(1,\dots,2l+1)$. The lowest form factor
  is $Q_{1,1}= e^{\frac{\theta'_1-\theta_1}{2}}$. We propose:
  \begin{eqnarray}
    \lefteqn{Q_{2r+1,2l+1}(\theta_1,\theta_2,\ldots
      \theta_{2r+1};\theta'_1,\theta'_2,\ldots,\theta'_{2l+1})=}
    \nonumber \\
    && (-i)^{r+l}\sum_{T \in S, \atop \#T=r}\sum_{T' \in S', \atop
      \#T'=l}\prod_{i \in T, \atop k\in
      \bar{T}}\phi(\theta_{ik})e^{\frac{1}{2}\sum\theta_{ik}}\prod_{i
      \in T',\atop  k\in \bar{T}'}\phi(\theta'_{ik})
    e^{\frac{1}{2}\sum\theta'_{ki}}\prod_{i \in T, \atop k\in
      \bar{T}'}\Phi(\theta_i-\theta'_k) \prod_{i \in T', \atop k\in
      \bar{T}}\tilde{\Phi}(\theta_k-\theta'_i) \nonumber
  \end{eqnarray}
\end{itemize}
Let us note that the exponentials will be responsible for the
additional minus sign in the residue condition (\ref{residu2}).
\subsubsection{Remarks}
\begin{itemize}
\item
  One can check that in the IR, the form factors of the operator
  $\mathcal{O}=\sigma+\mu$ satisfy the cluster property like an
  exponential of a bose field \cite{KM}, for example:
  $$
  \mathcal{O}_{r,l}(\theta_1,\theta_2,\dots,\theta_r;\theta'_1,\theta'_2,\dots,\theta'_l)
  \sim
  \mathcal{O}_{1,0}(\theta_1)\mathcal{O}_{r-1,l}(\theta_2,\dots,\theta_r;\theta'_1,\theta'_2,\dots,\theta'_l)
  \quad \text{for} \quad \theta_1 \to -\infty.
  $$
\item
  The expressions for the form factors of $\sigma$ and $\mu $ give
  the expected leading IR behaviour \cite{BKW,YZ,CM}: $
  F_{r,l}^{IR}(\theta_1,\theta_2,\ldots,\theta_r;\theta'_1,\theta'_2,\ldots,\theta'_l)
  \sim
  \prod_{i<j}\tanh\frac{\theta_{ij}}{2}\tanh\frac{\theta'_{ij}}{2}$,
  where $r+l$ is odd for $\sigma$ and even for $\mu$.
\end{itemize}

\section{Concluding remarks}
We understand it is important to check the UV properties of the
form factors proposed in this letter; we hope to present numerical
checks in a future publication. As far as the operators
$\Theta,\sigma,\mu$ are concerned, we expect our representation to
be the correct answer to the problem: indeed, the form factors of
the operators $\sigma$ and $\mu$ have the expected leading IR
behaviour; moreover we recover immediately the form factors of the
operators $\Theta, \sigma, \mu$ in the Tricritical Ising model
perturbed by the subenergy that defines a massless flow
to the Ising model \cite{KMS,AZ}, simply by replacing in our formulae
$S_{RL}$
and $f_{RL}$ by their
corresponding values that can be found in \cite{Z,DMS}. We checked
for a low number of particles that they correctly reproduce the
results of \cite{DMS} where the first form factors of the
operators $\Theta,\sigma,\mu$
are computed in terms of symmetric polynomials\footnote{The authors of \cite{DMS}
  checked numerically the UV properties of their form factors,
  including the c-theorem.}. We also obtained agreement (again for a
low number of particles) with \cite{MS}, where an expression quite
similar to ours for the form factors of the operator $\Theta$ is
proposed (with an arbitrary number of intermediate particles). The
case of the energy operator $\epsilon$ could be slightly more
tricky: although it is evoked in \cite{DMS}, only its
lowest form factor with one left mover and one right mover is
explicitly given there.\\
Finally, the representations we provide for the functions
$Q_{r,l}$ are in principle general enough\footnote{Their architecture
  is very similar to the one found for the form factors of the 
  operator $e^{\alpha\phi}$
  in the bosonic Sinh-Gordon model in \cite{BK}, equ.~(61) -let us recall that another representation in terms of determinant formula was first proposed in the prior work \cite{KM}.} to
provide results for other massless models flowing to the Ising
model,
but where the $S$-matrix has a more complicated structure of resonance poles \cite{Z}.

\section*{Acknowledgments}
I am grateful to Al.B.~Zamolodchikov for suggesting the problem.
Discussions with D.~Bernard, V.A.~Fateev, G.~Mussardo, H.~Saleur,
F.A.~Smirnov and especially G.~Delfino are acknowledged. Work
supported by the CEA and the Euclid Network HPRN-CT-2002-00325.

\end{document}